\documentclass[hyper]{JHEP} 

\usepackage{epsfig}

\newcommand\fverb{\setbox\pippobox=\hbox\bgroup\verb}
\newcommand\fverbdo{\egroup\medskip\noindent%
			\fbox{\unhbox\pippobox}\ }
\newcommand\fverbit{\egroup\item[\fbox{\unhbox\pippobox}]}
\newbox\pippobox
\title{Matrix model and Chern-Simons theory}

\author{by J. Kluso\v{n}\\
	 Department of Theoretical Physics and Astrophysics\\
                   Faculty of Science, Masaryk University\\
Kotl\'{a}\v{r}sk\'{a} 2, 611 37, Brno\\
Czech Republic\\
	E-mail: \email{klu@physics.muni.cz}}

\preprint{\hepth{0012184}}	

\abstract{In this short note we would like to present
a simple topological matrix model which has close 
relation with the noncommutative Chern-Simons theory.}


\def\tr{\mathrm{Tr}}

\begin{document}
\section{Introduction}
In recent years there was a great interest in the
area of the noncommutative theory and
its relation to string theory. In particular,
it was shown in the seminal paper \cite{WittenNG} 
 that the noncommutative theory
can be naturally embedded into the string theory.
It was also shown in the recent paper
\cite{Seiberg} that there is a remarkable 
connection between noncommutative gauge theories
and matrix theory. For that reason it is
natural to ask whether we can push
this correspondence further. In particular,
we would like to ask whether other gauge theories,
for example Chern-Simons theory, can be also
generalised to the case of noncommutative
ones. It was recently shown 
\cite{Polychronakos} that this can be done
in a relatively straightforward way in the case
of Chern-Simons theory. It is then natural
to ask whether, in analogy with 
\cite{Seiberg}, there is a relation between
topological matrix models \cite{Oda} and
Chern-Simons noncommutative theory.

It  was suggested in many papers \cite{Smolin1,
Smolin2,Horava} that Chern-Simons theory could
play profound  role in the nonperturbative
formulation of the string theory, M-theory. On the
other hand, one of the most successful (up
to date) formulation of M-theory is the Matrix
theory \cite{Banks}, for review, see
\cite{BanksR1,TaylorR1,BanksR2,TaylorR2,
SchwarzR}. We can  ask the question whether there
could be some connection between Matrix
models and Chern-Simons theory. This question
has been addressed in interesting papers
\cite{Smolin1,Smolin2}, where some very
intriguing ideas have been suggested. 

Noncommutative Chern-Simons theory could also
play an important role in the description of
the Quantum Hall Effect in the framework of string
theory \cite{Thomas,Susskind}. 
All these works suggest plausible possibility to
describe some configurations in the physics of
the condense systems in terms of
D-branes, which can be very promising area of
research. On the other hand,  some ideas of physics
of the condense systems could be useful 
in the nonperturbative 
formulation of the  string theory. In summary,
on all these examples we see that it is worth to
study the basic questions regarding to the
noncommutative
Chern-Simons theory and its relation to the matrix
theory and consequently to the string theory.
 
In this paper we will not address these exciting
ideas.  We will  rather ask the question whether
some from of the topological matrix model can lead to the
noncommutative Chern-Simons theory. Such a model has been
suggested in \cite{Smolin} and further elaborated in
\cite{Oda}.
We will show that the simple
 topological matrix model
\cite{Oda} cannot lead (As far as we know.) to
the noncommutative Chern-Simons theory. For that
reason we propose a simple modification of this model
when we include additional term containing the
information about the background structure of the
theory. Without including this term in the action
we would not be able to obtain noncommutative version
of the Chern-Simons theory. It is remarkable fact that
this term naturally arises in D-brane physics from
the generalised Chern-Simons term in D-brane
action in the presence of the background
Ramond-Ramond fields \cite{Mayers}. For that reason we believe that
our proposal could really be embedded in the string
theory and also could have some relation with 
M-theory.

\section{Brief review of Chern-Simons theory}
In this section we would like to review the basic
facts about Chern-Simons actions and in particular
their extensions to noncommutative manifolds. 
We will mainly follow \cite{Polychronakos}.

The Chern-Simons action is the integral of the $2n+1$ form
$C_{2n+1}$ over space-time manifold 
\footnote{In this article we will consider the Euclidean
space-times only.}which satisfies 
\begin{equation}
dC_{2n+1}=\tr F^{n+1} \ ,
\end{equation}
where the  wedge operation $\wedge$ between
forms $F$ is understood. The action
is defined as
\begin{equation}
\frac{\delta S_{2n+1}}{\delta A}=
\frac{\delta}{\delta A}\int C_{2n+1}=(n+1)F^n \ ,
\end{equation}
with the conventions
\begin{equation}
A=A_{\mu}dx^{\mu}, \ 
F=dA-iA\wedge A=\frac{1}{2}\left(\partial_{\mu}
A_{\nu}-\partial_{\nu}A_{\mu}-i[A_{\mu},A_{\nu}]\right)
dx^{\mu}\wedge dx^{\nu} \ .
\end{equation}
The extension of this action to the case of the noncommutative
background is straightforward
\cite{Polychronakos}. The easiest way to see this
is in terms of the  operator formalism 
 of the noncommutative geometry \cite{Alvarez}. 
In this short article we will not
discuss the operator formalism in more details since it
is well know from the literature (See \cite{Alvarez}
and reference therein.) 
The definition of the noncommutative Chern-Simons
action is very straightforward in the operator formalism
as was shown in \cite{Polychronakos}, where the whole
approach can be found. We will see the emergence of
the noncommutative Chern-Simons action in the operator
formalism in the next section where this action naturally
arises from the modified topological matrix model
\cite{Oda}.

\section{Matrix model of Chern-Simons theory}

It was argued in \cite{Oda,Smolin} that we can formulate
the topological matrix model which has many properties as 
 the Chern-Simons theory
\cite{Smolin}. The action for this model was proposed in
the form
\begin{equation}\label{actCS1}
S=\epsilon_{\mu_1\dots \mu_D}\tr X^{\mu_1}\dots X^{\mu_D} \ .
\end{equation}
 It is easy to see that this model can be defined in the odd dimensions only:
\begin{eqnarray}
S=\epsilon_{\mu_1\dots\mu_D}\tr X^{\mu_{D}}X^{\mu_1}
\dots X^{\mu_{D-1}}= \nonumber \\
=(-1)^{D-1}\epsilon_{\mu_D\mu_1
\dots \mu_{D-1}}\tr X^{\mu_D}X^{\mu_1}\dots X^{\mu_{D-1}}=
(-1)^{D-1}S \ , \nonumber \\
\end{eqnarray}
so we  have $D-1=2n\Rightarrow D=2n+1$.
The equations of motion obtained from (\ref{actCS1})
are
\begin{equation}
\frac{\delta S}{\delta X^{\mu}}=
\epsilon_{\mu\mu_1\dots \mu_D}X^{\mu_1}\dots
X^{\mu_D}=0, \ \mu=0,\dots,2n \ .
\end{equation}
It was argued in \cite{Oda} that there are solutions corresponding
to  D-branes. However, it is difficult to see whether
these solutions corresponding to some physical objects since
we do not know how to study the fluctuations around these solutions.
For example, for $D=3$ we obtain from the equation of motion
for $\mu=0,1,2$
\begin{equation}
[X^1,X^2]=0, \ [X^{0},X^1]=0\ , [X^2,X^0]=0 \ .
\end{equation}
We see that the only possible solutions correspond to separate objects 
where the matrices $X$ are diagonal or solution $X^1=0=X^2$ with
any $X^0$. We do not know any physical meaning of the second solution. For 
that reason we propose the modification of the topological matrix
model which, as we will see, has  a close relation with the 
noncommutative Chern-Simons theory
\cite{Polychronakos}. We propose the action in the form
\begin{equation}\label{actCS2}
S=(2\pi)^n\epsilon_{\mu_1\dots\mu_D}\tr
\left((-1)^{n/2}\frac{D+1}{2D}X^{\mu_1}\dots X^{\mu_D}+
(-1)^{(n-1)/2}\frac{D+1}{4(D-2)}\theta^{\mu_1\mu_2}
X^{\mu_3}\dots X^{\mu_D}\right) \ ,
\end{equation}
where $D=2n+1$ and the numerical factors $(-1)^{n/2}, (-1)^{(n-1)/2}$ arise
 from the requirement of
the reality of the action. The other factors  $(D+1)/(2D),
 \ (D+1)/(4(D-2))$ 
were introduced to have a contact with the work \cite{Polychronakos}.
In the previous expression (\ref{actCS2}) we have
also introduced the 
 matrix $\theta^{\mu\nu}$ which characterises given 
configuration.
The equations of motion have a form
\begin{equation}\label{EQ}
(-1)^{n/2}(n+1)\epsilon_{\mu\mu_1\dots \mu_{2n}}X^{\mu_1}
X^{\mu_2}\dots X^{\mu_{2n}}+(-1)^{(n-1)/2}\frac{n+1}{2}\epsilon
_{\mu\mu_1\mu_2\dots \mu_{2n}}\theta^{\mu_1\mu_2}
X^{\mu_3}\dots X^{\mu_{2n}}=0 \ .
\end{equation}
 We would like to find  solution corresponding to
 the noncommutative Chern-Simons action.
From the fact that we have odd number of dimensions we
see that one dimension should correspond to the commutative
one. In order to obtain Chern-Simons action in the
noncommutative space-time  we
will follow \cite{TaylorTT} and compactify the commutative
direction $X^0$.
 For that  reason we write any matrix as
\begin{equation}
X^{\mu}_{IJ}=(X^{\mu}_{ij})_{mn}, \
I=m\times M+i, \ J=n\times M+j , \ 
\end{equation}
where  $X_{ij}$ is $M\times M $ matrix
with $M\rightarrow \infty $ and also $m,n$ go  from
$-N/2$ to $N/2$ and  we again take the limit $N\rightarrow \infty$.
In other words, the previous expression corresponds to
the direct product of the matrices
\begin{equation}
X=A\otimes B \Rightarrow X_{xy}=A_{ij}B_{kl} , \ x=i\times M+k, \
y=j\times M+l , 
\end{equation}
with  $M\times M$ matrix $B$. We
impose the following constraints on the various matrices
\cite{TaylorTT}
\begin{eqnarray}\label{twist}
(X^{i}_{ij})_{mn}=(X^{i}_{ij})_{m-1,n-1}, \
a=1,\dots, 2n \ , \nonumber \\
(X^0_{ij})_{mn}=(X^{0}_{ij})_{m-1,n-1} \ , m\neq n \ , \nonumber \\
(X^{0}_{ij})_{nn}=2\pi R\delta_{ij}+(X^{0}_{ij})_{n-1,n-1} \ ,
\nonumber \\
\end{eqnarray}
where $R$ is a radius of compact dimension. 
These constraints (\ref{twist}) can be solved as
\cite{TaylorTT}
\begin{eqnarray}
(X^i_{ij})_{mn}=(X^i_{ij})_{0,m-n}=
(X^i_{ij})_{m-n} \ , \nonumber \\
(X^{0}_{ij})_{mn}=2\pi R m \delta_{mn}\otimes
\delta_{ij}+
(X^0_{ij})_{m-n} \ . \nonumber \\
\end{eqnarray}
   We then immediately obtain
\begin{eqnarray}\label{pom}
([X^0,X^i]_{ij})_{mp}=2\pi Rm\delta_{mn}(X^i_{ij})_{np}-
(X^i_{ij})_{mn}2\pi R n\delta_{np}+\nonumber \\
+(X^0_{ik})_{mn}
(X^i_{kj})_{np}-(X^i_{ik})_{mn}(X^0_{kj})_{np}=\nonumber \\
=2\pi R (m-p)(X^i_{ij})_{m-p}+([X^0,X^i]_{ij})_{m-p} \ ,\nonumber \\
\end{eqnarray}
where $(X^{\mu}_{ij})_{0m}=(X^{\mu}_{ij})_m$.
We see that the commutator $X^0$ with any $X^i$ has a form 
of the covariant derivative \cite{TaylorTT} where the first term correspond
to the ordinary derivative $-i\partial_0 $ with respect to the
dual coordinate $\tilde{x}_0$ which is identified 
as $\tilde{x}_0\sim x_0+2\pi /R$. The second term
is  the commutator of the gauge field $X^0=A_0$ with any matrix. We could then
proceed as in \cite{TaylorTT} and rewrite the action in
the form of the dual theory defined on the dual torus with
the radius $\tilde{R}=1/R$, but for simplicity we will use
the original variables. Using
this result we will write
$X^0=KC_0$ as 
\begin{equation}
(C_{0,ij})_{mn}=p_{0,mn}\otimes 1_{M\times M}+(A_{0,ij})_{mn} \ ,
\end{equation}
where the acting of $p_0$ on various matrices
is defined in  (\ref{pom}) and where the numerical factor
$K$ will be determined for letter convenience.

For  illustration of the main idea,
 let us consider matrix model defined in
 $D=2n+1=3$ dimensions. Let us consider
the matrix $\theta^{\mu\nu}$ in the form
\begin{equation}
\theta^{\mu\nu}=1_{mn}\otimes\left(\begin{array}{ccc}
0 & 0 & 0 \\
0 & 0 & \theta 1_{N\times N}\\
0 & -\theta 1_{N\times N} & 0 \\ \end{array}\right) \ ,
\end{equation}
where $1_{N\times N}$ is a unit matrix  with $N$ going to 
infinity. Then the equations of motion, which arise
from (\ref{actCS2}), are
\begin{eqnarray}\label{eq2}
i\epsilon_{0 12}X^{1}X^{2}+i\epsilon_{021}
X^2X^1+\frac{1}{2}\left(
\epsilon_{012}\theta^{12}+\epsilon_{021}\theta^{21}\right)=0
\nonumber \ , \\  
i\epsilon_{102}X^0X^2+i\epsilon_{120}X^2X^0=0 \ , \nonumber \\ 
i\epsilon_{201}X^0X^1+i\epsilon_{210}X^1X^0=0 \ , \nonumber \\ 
\end{eqnarray}
The second and the third equation gives the condition $[X^0,X^i]=0$ which
leads to the solution $A_0=0$ and $[p_0,X^i]=0$. These equations, together
with the first one, 
can be solved as
\begin{equation}\label{clas}
X^i=\delta_{mn}\otimes x^i_{jk} \ ,[x^i,x^j]=i\theta^{ij} \ .
\end{equation}
Thanks to the presence of the unit matrix $\delta_{mn}$, $X^i$
 commutes with $p_0\otimes 1_{M\times M}$ and so is 
the solution of the equation of motion (\ref{eq2}).
 Following \cite{Seiberg}, we can study the fluctuations
around this solution  with using the ansatz
\begin{eqnarray}\label{ans}
X^0=\omega_{12}C_0=\omega_{12}(p_0\otimes 1_{M\times M}+
(A_{0, ij})_{mn}) , \ \nonumber \\
X^i=\theta^{ij}C_j, 
\ C_i=1_{N\times N}\otimes p_i+(A_{i,ij})_{mn}, 
\ p_i=\omega_{ij}x^j, \ i=1,2 \ ,
\omega_{ij}=(\theta^{-1})_{ij} \ , \nonumber \\
\end{eqnarray}
It is easy to see that this  configuration corresponds 
to the noncommutative Chern-Simons action in $D=3$ dimensions
 \cite{Polychronakos}. More precisely, let us
 introduce formal
parameters
\begin{equation}
dx^{\mu} , \ \mu=0,\dots, 2, \  dx^{\mu}\wedge dx^{\nu}=
-dx^{\nu}\wedge dx^{\mu}, \epsilon^{\mu_{1}
\dots \mu_{D}}=dx^{\mu_1}\wedge \dots
\wedge dx^{\mu_D} 
\end{equation}
and the matrix valued one form
\begin{equation}
C=C_{\mu}dx^{\mu}=d+A \ ,
\end{equation}
where $C_{\mu}$ is given in (\ref{ans}).
Then  the action describing the fluctuations around
the classical solution (\ref{clas}) has a form
\begin{equation}\label{Css3}
S=2\pi\sqrt{\det \theta}\tr\left(
-i\frac{2}{3}  C\wedge C \wedge C 
+2  \omega\wedge C\right) \ .
\end{equation}
We rewrite this action in the form which has a closer contact with
 the commutative Chern-Simons theory. 
Firstly we prove the cyclic symmetry of the trace of the forms
\begin{eqnarray}
\tr \left(A^1\wedge \dots A^D\right)=
\tr \left(A^1_{\mu_1}A^2_{\mu_2}\dots A^D_{\mu_D}\right)
dx^{\mu_1}\wedge \dots \wedge dx^{\mu_D}=\nonumber \\
=\tr \left(A^2_{\mu_2}\dots A^D_{\mu_D}A^1_{\mu_1}\right)
dx^{\mu_2}\wedge dx^{\mu_D}\wedge dx^{\mu_1}=
\tr \left(A^2\wedge \dots\wedge A^D\wedge A^1\right) \ , \nonumber \\
\end{eqnarray}
where we have used the fact that
$D$ is odd number so that  $dx^{\mu_1}$ commutes
with even numbers of $dx$. Then the expression (\ref{Css3})
is equal to
\begin{equation}
S=-2\pi\sqrt{\det \theta}\tr\left(
iA\wedge (d\wedge A+A\wedge d)+\frac{i2}{3} A\wedge A\wedge
A\right) \ ,
\end{equation}
where we have used 
\begin{eqnarray}\label{pp}
d\wedge d=p_{\mu}p_{\nu}dx^{\mu}dx^{\nu}=
\frac{1}{2}[p_{\mu},p_{\nu}]dx^{\mu}\wedge dx^{\nu}=
-i \omega , \nonumber \\
 \omega=\frac{1}{2}\omega_{\mu\nu}dx^{\mu} 
\wedge dx^{\nu}, \ \omega_{ij}=(\theta^{-1})_{ij}, \ 
\omega_{0i}=0 \ . \nonumber \\
\end{eqnarray}
Now it is easy to see that (\ref{Css3}) is a
correct action for the fluctuation fields  $A$. 
The equations of
motion arising from (\ref{Css3}) are
\begin{equation}
-2i (d+A)\wedge (d+A)+2\omega =0 \ .
\end{equation}
Looking at (\ref{pp}) it is easy to see that the configuration $A=0$
is a solution of equation of motion as it should be for
the fluctuating field.
With using
\begin{equation}
d\wedge A+A\wedge d=[p_{\mu},A_{\nu}]dx^{\mu}
\wedge dx^{\nu}=i\partial_{\mu} A_{\nu}dx^{\mu}\wedge dx^{\nu}
=i d\cdot A \ ,
\end{equation}
we obtain the  derivative $d\cdot$ 
that is an analogue of the exterior derivative in the ordinary commutative
geometry. In this case the action has a form
\begin{equation}\label{CS3}
S=2\pi\sqrt{\det \theta}\tr\left(A\wedge d\cdot A-
\frac{2i}{3}A\wedge A\wedge A\right) \ ,
\end{equation}
which  is the standard Chern-Simons action in three dimensions.

 We
observe that this action differs from the action given in \cite{Polychronakos}
since there is no the term $\omega\wedge A$ in our action. 
This is a consequence of the presence of the second term in (\ref{actCS2}) that is needed for
the emergence of noncommutative structure in the Chern-Simons action.
On the other hand, from the fact that similar matrix
structure arises in the study of Quantum Hall Effect in D-brane
physics \cite{Susskind} we believe that our proposal
of topological action could have relation to the string
theory and M theory.
 As usual,
this action can be rewritten using in terms of the integral
over space-time with ordinary multiplication replaced with
star product \cite{Alvarez}.

Generalisation to the higher dimensions is straightforward. The equations
of motion (\ref{EQ}) give
\begin{equation}
i\epsilon_{\mu\nu}X^{\mu}X^{\nu}+
\frac{1}{2}\epsilon_{\mu\nu}\theta^{\mu\nu}=0
\Rightarrow
[X^{\mu},X^{\nu}]=i\theta^{\mu\nu}, \ \mu, \ \nu=
1,\dots, 2n \ .
\end{equation}
We restrict ourselves to the case of $\theta$ of
 the maximal rank.
For simplicity, we consider $\theta$ in the form
\begin{equation}
\theta^{\mu\nu}=\left(
\begin{array}{cccccc}
0 & \dots & \dots & \dots & \dots & 0\\
0 & 0 & \theta_1 & 0  & \dots & 0 \\
0 & -\theta_1 & 0 & \dots & \dots & 0 \\
\dots & \dots & \dots & \dots & \dots &\dots \\
0 & \dots & \dots & \dots & 0 & \theta_n \\
0 & \dots & \dots & 0 & -\theta_n & 0 \\
\end{array}\right) \ .
\end{equation}
As in $3$ dimensional case we introduce the matrix $\omega$
defined as follows
\begin{equation}
\omega_{ij}=(\theta^{-1})_{ij} , \ i,j=1,\dots, 2n, \
\omega_{i0}=\omega_{0i}=0 \ .
\end{equation}
and we define $C=C_{\mu}dx^{\mu}=
d+A \ , \mu=0, \dots, 2n$ where the dimension $x^0$ 
is compactified as above. 
And finally various $C_{\mu}$ are defined as 
\begin{equation}
X^i=\theta^{ij}C_j, \ X^0=\left(\prod_{i=1}^n \omega_i\right) C_0 \ ,
\omega_i=-\theta^{-1}_i \ ,
\end{equation}
with as $C_{\mu}$   same as in (\ref{ans}).
 Then the action has   a form
\begin{eqnarray}
S_{2n+1}=(2\pi)^n\sqrt{\det \theta}
\tr \left((-1)^n(-1)^{n/2}\frac{n+1}{2n+1}C^{2n+1}+\right.\nonumber \\
\left.+(-1)^{n-1}
(-1)^{(n-1)/2}\frac{n+1}{2n-1} \omega 
\wedge C^{2n-1}\right) \ . \nonumber \\
\end{eqnarray}
In order to obtain more detailed description of
the action we will follow \cite{Polychronakos}.
Since $\frac{\delta }{\delta C}=\frac{\delta}{\delta A}$,
we can write
\footnote{We will write $C^n$ instead of $C\wedge \dots 
\wedge C$.}
\begin{eqnarray}\label{delta}
\frac{\delta S_{2n+1}}{\delta C}=(2\pi)^n
\sqrt{\det \theta}\left((-1)^n
(-1)^{n/2}(n+1) C^{2n}+\right. \nonumber \\
\left. +(-1)^{n-1}(-1)^{(n-1)/2}(n+1)
\omega \wedge C^{2n-2}\right)=\nonumber \\
=(2\pi)^n\sqrt{\det\theta}\left(
 (n+1)(F-\omega)^n+(n+1)\omega \wedge
(F-\omega )^{n-1} \right)\ , \nonumber \\
 \end{eqnarray}
where we have used the fact  that 
$C^2=-i\omega+i(-idA-iAd-iA^2)=-i\omega+iF$.
Since $\omega$ and $F$  are both  two forms and
 $\omega$ is a pure number from the point of view 
 of the  noncommutative
geometry we immediately see that $F$ and $\omega$
commute so that we can write
\begin{equation}
(F-\omega)^n=\sum_{k=0}^n\left(
\begin{array}{cc} n \\
                          k \\ \end{array}\right)
(-\omega)^{n-k}F^k \ .
\end{equation}
 Following \cite{Polychronakos} we introduce the other
form of the Lagrangian
\begin{equation}
\frac{\delta \tilde{L}_{2k+1}}{\delta C}=
(k+1) F^k  \ .
\end{equation}
Then we can rewrite (\ref{delta}) as
\begin{eqnarray}\label{Agen}
\frac{\delta}{\delta C}\left\{
S_{2n+1}-(2\pi)^{n}\sqrt{\det\theta}
\left(\sum_{k=0}^n
\left(
\begin{array}{cc} n+1 \\
                          k+1 \\ \end{array}\right)
(-\omega)^{n-k}\tilde{L}_{2k+1}\right. \right. \nonumber \\
\left.\left. -\sum_{k=0}^{n-1}
(-1)^{n-k-1}\frac{1}{n}
\left(\begin{array}{cc} n+1 \\
                          k+1 \\ \end{array}\right)
\omega^{n-k} \wedge 
\tilde{L}_{2k+1}
\right)\right \}=0 \Rightarrow\nonumber \\
\Rightarrow S_{2n+1}=(2\pi)^n
\sqrt{\det\theta}\tr\left(\sum_{k=0}^n
(-1)^{n-k}
\left(\begin{array}{cc} n+1 \\
                          k+1 \\ \end{array}\right)
\omega^{n-k}\tilde{L}_{2k+1}\right.  +\nonumber \\
\left. + \sum_{k=0}^{n-1}
(-1)^{n-k-1}\frac{1}{n}
\left(\begin{array}{cc} n+1 \\
                          k+1 \\ \end{array}\right)
\omega^{n-k} \wedge 
\tilde{L}_{2k+1}\right) \ .
 \nonumber \\
\end{eqnarray}
As a check, for $n=1$ we obtain from (\ref{Agen})
\begin{equation}
S_3=(2\pi)\sqrt{\det \theta}\tr\left(
-2\omega\wedge \tilde{L}_1+\tilde{L}_3+2\omega
\wedge \tilde{L}_1\right)=
2\pi\sqrt{\det \theta}\tr \tilde{L}_3 \ ,
\end{equation}
and using
\begin{equation}
\frac{\delta \tilde{L}_3}{\delta A}
=2F=-2i(d A+Ad+A^2)\Rightarrow
\tilde{L}_3=-iA\wedge d\wedge A
-id\wedge A\wedge A-i\frac{2}{3}A\wedge A
\wedge A \ ,
\end{equation}
we obtain
\begin{equation}
S_3=2\pi\sqrt{\det \theta}
\tr \left( A\wedge d\cdot A-i\frac{2}{3}
A\wedge A\wedge A \right) \ ,
\end{equation}
which, as we have seen above, is a correct form
of the noncommutative Chern-Simons action
in three dimensions. 

\section{Conclusion}
In this short note we have shown that simple
modification of the topological matrix model
\cite{Oda} could lead to the emergence of the noncommutative
Chern-Simons action \cite{Polychronakos}. 
In order to obtain this action we had to introduce the 
antisymmetric matrix $\theta $ expressing the noncommutative
nature of the space-time. It is crucial fact that
we must introduce this term into the action explicitly
which differs from the case of the standard matrix
theory \cite{Seiberg}, where different configurations
with any values of the noncommutative parameters
arise as particular solutions of the matrix theory.  

It is also clear that we can find much more configurations
than we have shown above. The form of these configurations
depend on $\omega$. It is possible to find such a $\theta$
which leads to the emergence of lower dimensional 
Chern-Simons actions and also which leads to the emergence
of point-like degrees of freedom in the Chern-Simons theory.
For example, we can consider $\theta $ in the form
\begin{equation}
\theta^{\mu\nu}=1_{mn}\otimes\left(\begin{array}{ccc}
0 & 0 & 0 \\
0 & 0 & A \\
0 & -A & 0 \\ \end{array}\right) \ ;
A=\left(\begin{array}{cc} 
0_{k\times k} & 0 \\
0 & \theta 1_{N\times N} \\ \end{array}\right) \ .
\end{equation}
This corresponds to the configuration describing Chern-Simons
action  with the presence of $k$
 point-like degrees of freedom - "partons". We could analyse the
interaction between these partons and gauge fields in the
same way as in matrix theory
(For more details see \cite{TaylorR1,TaylorR2} and reference
therein.) It is possible that this simple model could have
some relation to the holographic model of M-theory
\cite{Horava}. In particular, we see that 
the partons arise naturally in  our approach. On the other hand,
the similar analysis as in \cite{Horava} could determine
$\theta$, i.e. Requirements of the consistency of 
the theory could choose $\theta$ in some particular form.
In short, we hope
 that the approach given in this paper could 
shine some light on the relation between the matrix
models and Chern-Simons theory. 
\\
\\
{\bf Acknowledgements}
We would like to thank Rikard von Unge
for many helpful discussions.
 This work was supported by the
Czech Ministry of Education under Contract No.
144310006.

\end{document}